\documentclass{article}
\usepackage{amsmath, amssymb, amsfonts}
\usepackage{float}

\usepackage{authblk}  % <-- THIS IS CRITICAL

\usepackage{graphicx}
\usepackage{hyperref}
\usepackage{PRIMEarxiv}
\usepackage[utf8]{inputenc} % allow utf-8 input
\usepackage[T1]{fontenc}    % use 8-bit T1 fonts
\usepackage{hyperref}       % hyperlinks
\usepackage{url}            % simple URL typesetting
\usepackage{booktabs}       % professional-quality tables
\usepackage{amsfonts}       % blackboard math symbols
\usepackage{nicefrac}       % compact symbols for 1/2, etc.
\usepackage{microtype}      % microtypography
\usepackage{lipsum}
\usepackage{fancyhdr}       % header
\usepackage{graphicx}       % graphics
\graphicspath{{media/}}     % organize your images and other figures under media/ folder

\title{MedFormer-UR: Uncertainty-Routed Transformer for Medical Image Classification
}

\author{
  Mohammed Maaz Sibhai \thanks{[1] 
 Department of Computer Science, Lakehead University} \\
  \texttt{msibhai@lakeheadu.ca}
  \and
  Abedalrehman Alkhateeb \\
  \texttt{aalkhate@lakeheadu.ca}
  \and
  Saad B. Ahmed \\
  \texttt{sbinahm@lakeheadu.ca}
}
\begin{document}
\maketitle

\begin{abstract}
To ensure safe clinical integration, deep learning models must provide more than just high accuracy; they require dependable uncertainty quantification. While current Medical Vision Transformers perform well, they frequently struggle with overconfident predictions and a lack of transparency, issues that are magnified by the noisy and imbalanced nature of clinical data.
To address this, we enhanced the modified Medical Transformer (MedFormer) that incorporates prototype-based learning and uncertainty-guided routing, by utilizing a Dirichlet distribution for per-token evidential uncertainty, our framework can quantify and localize ambiguity in real-time. This uncertainty is not just an output but an active participant in the training process, filtering out unreliable feature updates. Furthermore, the use of class-specific prototypes ensures the embedding space remains structured, allowing for decisions based on visual similarity. Testing across four modalities (mammography, ultrasound, MRI, and histopathology) confirms that our approach significantly enhances model calibration, reducing expected calibration error (ECE) by up to 35\%, and improves selective prediction, even when accuracy gains are modest.

\end{abstract}

% keywords can be removed


\section{Introduction}
Deep learning has achieved remarkable success in medical image analysis, with vision Transformers (ViTs) emerging as powerful alternatives to convolutional neural networks (CNNs) due to their ability to capture long-range dependencies via self-attention \cite{vaswani2017transformer, shamshad2023transformersurvey}. In medical imaging, hierarchical vision Transformers such as MedFormer \cite{xia2025medformer} have demonstrated state-of-the-art performance across segmentation and classification tasks.

However, the clinical deployment of these models faces a critical challenge: \textit{trustworthiness}. Clinicians require not only accurate predictions but also reliable uncertainty estimates and interpretable rationales. Standard deep learning models produce poorly calibrated probabilities, often exhibiting overconfidence on out-of-distribution or ambiguous inputs \cite{guo2017calibration}. Moreover, Transformer-based models operate as black boxes, offering little insight into \textit{why} a particular decision was made.

Several approaches have attempted to address these limitations. Bayesian methods, such as Monte Carlo Dropout \cite{gal2016dropout}, provide uncertainty estimates but incur substantial computational overhead during inference. Post-hoc calibration techniques, most notably Temperature Scaling \cite{guo2017calibration}, improve probability alignment but do not mitigate overconfidence during the training phase. Furthermore, common explainability methods like attention visualization offer qualitative interpretability but often remain disconnected from the model's actual decision-making process \cite{rudin2019interpretable}.
This paper introduces \textbf{MedFormer-UR} Transformer with Uncertainty Routing based-on Prototype Learning for Medical Image Classification, a unified framework that integrates:
\begin{enumerate}
    \item \textit{Evidential uncertainty estimation} using a Dirichlet distribution, providing per-token uncertainty quantification without sampling,
    \item \textit{Uncertainty-guided feature routing} where high-uncertainty tokens are dynamically suppressed or refined during both forward and backward passes, and
    \item \textit{Prototype-based learning} that maintains class-specific embedding prototypes, enabling similarity-based reasoning and interpretable predictions.
\end{enumerate}

Our key contributions are:
\begin{itemize}
    \item A novel extension of MedFormer with evidential heads that produce Dirichlet parameters, enabling explicit aleatoric and epistemic uncertainty decomposition.
    \item Uncertainty-guided routing mechanisms that modulate token contributions based on estimated uncertainty, reducing the influence of ambiguous features during training.
    \item Integration of learnable class prototypes with a contrastive prototype loss, structuring the embedding space for interpretable similarity-based classification.
    \item Comprehensive evaluation on four medical imaging datasets spanning mammography, ultrasound, MRI, and histopathology, demonstrating consistent improvements in calibration (18--35\% ECE reduction) and selective prediction capabilities.
\end{itemize}

\section{Datasets}

This chapter describes the datasets used to evaluate the proposed uncertainty-aware MedFormer across multiple medical imaging modalities. The selected datasets span radiological and histopathological imaging and include both binary and multi-class classification tasks. Together, they provide a diverse and challenging benchmark for assessing model performance, robustness, and uncertainty behaviour. For each dataset, the imaging modality, classification task, dataset size, image characteristics, and relevant challenges are described. All datasets used in this research are publicly available.

\subsection{CBIS-DDSM (Mammography)}

The Curated Breast Imaging Subset of the Digital Database for Screening Mammography (CBIS-DDSM) is a curated and standardized version of the original DDSM dataset, created to facilitate reproducible breast cancer research \cite{lee2017cbisddsm}. It includes digitized film mammograms acquired during routine screening, with expert-annotated findings and pathology-confirmed labels. CBIS-DDSM contains images of two primary abnormality types; masses and calcifications, each annotated as benign or malignant. In this study, the classification task is formulated as a binary problem, distinguishing benign from malignant cases, where benign-without-callback annotations are grouped with benign cases, following standard CBIS-DDSM practice.

In this study, the classification task is formulated as a binary problem that distinguishes benign from malignant cases. Benign-without-callback annotations are grouped with benign cases, following common practice in CBIS-DDSM-based studies. %As illustrated in Figure~\ref{fig:cbis}, 
The dataset exhibits a moderate class imbalance after merging categories. Both craniocaudal (CC) and mediolateral oblique (MLO) views are included, reflecting standard clinical screening protocols. Overall, the dataset contains over 3,000 grayscale mammograms from more than 1,500 studies, often with very high spatial resolutions exceeding $3000 \times 2000$ pixels. Due to the large image sizes, mammograms are typically resized or processed using patch-based strategies prior to model training.

%\begin{figure}[htbp]
%\centering
%\includegraphics[width=0.65\textwidth]{CBIS.png}
%\caption{Class distribution of the CBIS-DDSM full mammogram dataset. The dataset contains 40.82\% malignant cases, 40.04\% benign cases, and 19.12\% benign-without-callback cases.}
%\label{fig:cbis}
%\end{figure}

\subsection{BUSI (Breast Ultrasound Images)}

The BUSI dataset is a publicly available collection of breast ultrasound images created to facilitate research on automated lesion classification \cite{aldhabyani2020busi}. Ultrasound imaging is commonly used alongside mammography in clinical settings, especially for patients with dense breast tissue. The dataset is labelled into three categories: benign lesions, malignant lesions, and normal breast tissue. Therefore, this study treats the task as a three-class classification problem. BUSI contains 780 ultrasound images from 600 patients. These images are in grayscale, typically 500 × 500 pixels, though sizes can vary. The dataset also provides pixel-level lesion segmentation masks, but this work only uses the image-level class labels. %Figure~\ref{fig:busi} presents the class distribution across the three diagnostic categories, highlighting a moderate class imbalance.

%\begin{figure}[htbp]
%\centering
%\includegraphics[width=0.65\textwidth]{BUSI.png}
%\caption{Class distribution of the BUSI breast ultrasound dataset across benign, malignant, and normal categories.}
%\label{fig:busi}
%\end{figure}

\subsection{Breast Histopathology (IDC)}

The breast histopathology dataset aims to identify invasive ductal carcinoma (IDC) in microscopic tissue images \cite{spanhol2015breakhis}. Histopathology is considered the gold standard for definitive cancer diagnosis, offering cellular-level detail through stained tissue sections. The dataset includes image patches taken from whole-slide images of breast biopsy samples. Each patch is labeled as either IDC or non-IDC, forming a binary classification task. Overall, the dataset contains about 277,000 image patches, derived from 162 whole-slide images from 82 patients. Each patch measures 50 × 50 pixels and is stored as an RGB image, reflecting the typical hematoxylin-and-eosin (H\&E) staining. %The class distribution for IDC and non-IDC samples is shown in Figure~\ref{fig:breakhis}.

%\begin{figure}[htbp]
%\centering
%\includegraphics[width=0.65\textwidth]{BreakHis.png}
%\caption{Class distribution of the BreaKHis histopathology dataset for invasive ductal carcinoma (IDC) classification.}
%\label{fig:breakhis}
%\end{figure}

\subsection{Brain MRI (Four-Class Tumor Classification)}

The brain MRI dataset employed in this study is designed for multi-class brain tumor classification and has been extensively utilized in medical image analysis studies \cite{gomezguzman2025braintumor}. MRI offers excellent soft-tissue contrast, making it a routinely preferred method for diagnosing and evaluating brain tumors. %The dataset encompasses four tumor categories, each representing different brain abnormalities as shown in Figure ~\ref{fig:btmri}.
The classification task is set as a four-class problem, where the model must distinguish tumor types based on MRI slice appearance. It contains 7,023 MRI images, with 5,618 allocated for training and 1,405 for testing. Images are typically resized to a uniform 224 × 224 pixels and may originate from various MRI scanners and protocols. The brain MRI dataset employed in this study is designed for multi-class brain tumor classification and has been extensively utilized in medical image analysis studies \cite{gomezguzman2025braintumor}. 

The classification task is defined as a four-class problem in which the model must distinguish tumor types based on MRI slice appearance. The dataset contains 7,023 MRI images, with 5,618 images allocated for training and 1,405 for testing. Images are resized to a uniform resolution of $224 \times 224$ pixels and originate from multiple scanners and imaging protocols. For this study, the dataset is used in its publicly available form \cite{abouali2021brainmri}. %Figure~\ref{fig:btmri} shows the class distribution across the four categories.

%\begin{figure}[htbp]
%\centering
%\includegraphics[width=0.65\textwidth]{BTMRI.png}
%\caption{Class distribution of the brain tumor MRI dataset used in this study, comprising glioma, no tumor, pituitary tumor, and meningioma categories.}
%\label{fig:btmri}
%\end{figure}

\subsection{Summary of Datasets}

The datasets used in this study span multiple imaging modalities and classification scenarios, enabling a comprehensive evaluation of model performance, calibration, and robustness. By covering both binary and multi-class tasks across radiological and histopathological data, the experimental analysis is grounded in diverse clinical settings.
\renewcommand{\arraystretch}{1.3}
\begin{table}[htbp]
\centering
\caption{List of datasets used for model performance evaluation}
\footnotesize
\begin{tabular}{l l l c}
\hline
Dataset & Modality & Task Type & Classes \\
\hline
CBIS-DDSM & Mammography (X-ray) & Binary & 2 \\
BUSI & Ultrasound & Multi-class & 3 \\
Histopathology & Microscopy & Binary & 2 \\
Brain MRI & MRI & Multi-class & 4 \\
\hline
\end{tabular}
\label{tab:datasets}
\end{table}

Together, these datasets ensure that the proposed methods are evaluated across a wide range of visual characteristics, noise sources, and clinical challenges, strengthening the generalisability and credibility of the experimental results.

% =========================
% Chapter 4 — Methodology
% =========================

\section{Methodology}

This chapter outlines the methodological foundation of the proposed approach and serves as its main contribution. While the baseline MedFormer architecture provides a strong hierarchical transformer backbone for medical image classification, it remains fundamentally deterministic and lacks mechanisms for understanding uncertainty, selectively managing feature propagation, or providing interpretable decision rationales. These limitations are especially important in medical imaging, where ambiguous tissue appearance, inter-patient variability, and distribution shifts are common.

To address these issues, this work presents an uncertainty-aware and prototype-augmented extension of MedFormer. The new architecture incorporates per-token evidential uncertainty estimation, uncertainty-guided feature routing and refinement, and prototype-based decision making into the MedFormer backbone, without modifying its core Dual Sparse Selection Attention (DSSA) mechanism. The resulting model strives to be both accurate and trustworthy by explicitly modeling uncertainty and producing structured, interpretable representations.

% -------------------------
% 4.1 Baseline MedFormer
% -------------------------
\section{Baseline Architecture: Original MedFormer}

\begin{figure}[H]
\centering
\includegraphics[width=0.75\textwidth]{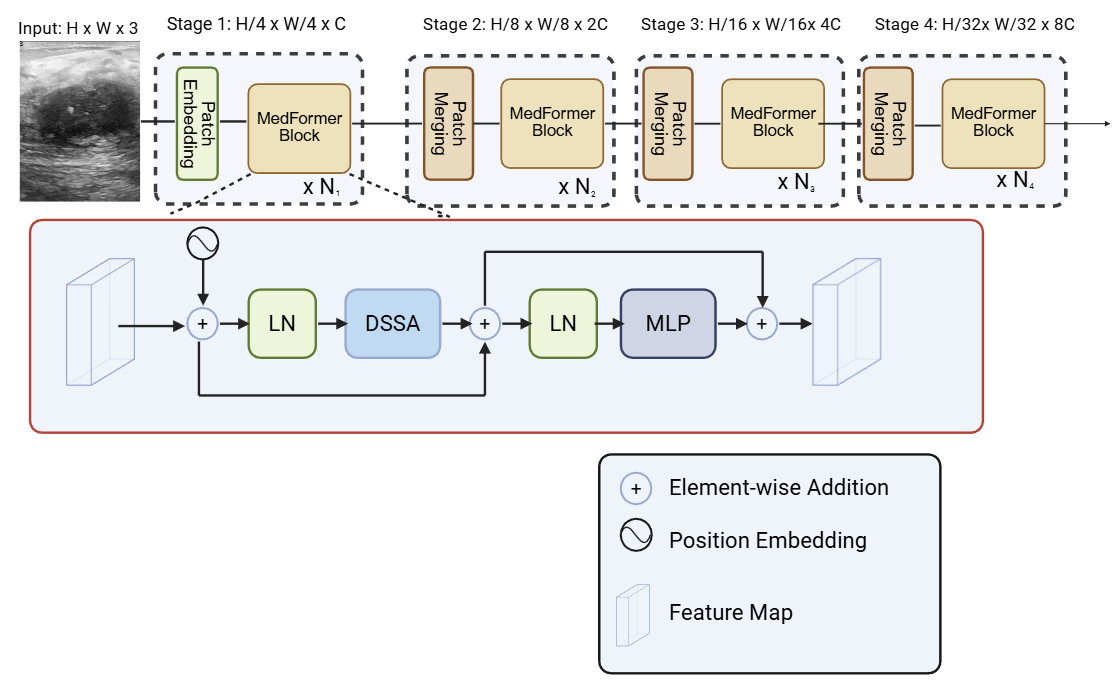}
\caption{Baseline MedFormer architecture with DSSA attention.}
\label{fig:medformer}
\end{figure}

MedFormer is a hierarchical vision Transformer designed to capture both detailed local structures and high-level semantic context in high-resolution images. The architecture comprises four stages (shown in Figure~\ref{fig:medformer}) that operate at progressively lower spatial resolutions and higher channel dimensionality, enabling efficient modelling of large medical images such as mammograms, histopathology slides, ultrasound scans, and MRI slices.

Patch embedding in MedFormer is performed using a convolutional stem consisting of two strided convolutions. Unlike standard Vision Transformers, which tokenize images with fixed-size, non-overlapping patches, this approach maintains local spatial continuity and boundary information, which is essential for medical images where subtle texture differences and lesion edges are important for diagnosis.

Within each stage, MedFormer employs Dual Sparse Selection Attention (DSSA). DSSA limits attention computation to a subset of spatial tokens selected via a top-down approach, enabling the model to concentrate computational resources on informative regions while keeping the process manageable for high-resolution inputs. This attention mechanism is followed by a feed-forward MLP and residual connections, forming a standard Transformer block structure adapted for dense medical imagery.

After the final stage, the spatial feature map $\mathbf{F}$ is normalized and aggregated using global average pooling to produce a compact image-level representation $\mathbf{f}\in\mathbb{R}^{D}$:
\begin{equation}
\mathbf{f}=\frac{1}{HW}\sum_{i=1}^{H}\sum_{j=1}^{W}\mathbf{F}_{i,j}.
\label{eq:gap}
\end{equation}
Here, $\mathbf{F}\in\mathbb{R}^{H\times W\times D}$ denotes the output feature map of the final MedFormer stage, where $H$ and $W$ are the spatial dimensions of the feature grid and $D$ is the channel (embedding) dimension. Each spatial location $(i,j)$ corresponds to a feature vector $\mathbf{F}_{i,j}\in\mathbb{R}^{D}$. Normalization is applied along the channel dimension prior to pooling.

The pooled representation is fed into a linear classifier and trained using standard cross-entropy loss. Although this baseline performs well in terms of accuracy, it often yields point estimates that are miscalibrated and overly confident, motivating the proposed extensions.

% -------------------------
% 4.2 Proposed Method
% -------------------------
\section{Uncertainty and Prototype Augmented MedFormer}

The TRU-MedFormer architecture is inspired by evidential uncertainty modelling, uncertainty-aware learning, sparse attention mechanisms, and prototype-based representation learning. Although these ideas have been studied separately, this study presents a unified framework that combines them within a hierarchical medical Transformer and uses uncertainty as an active control signal for feature refinement and routing. The main innovations include the architectural design, the training objectives, and how uncertainty-guided routing interacts with prototype-based regularization.

\begin{figure}[H]
\centering
\includegraphics[width=0.75\textwidth]{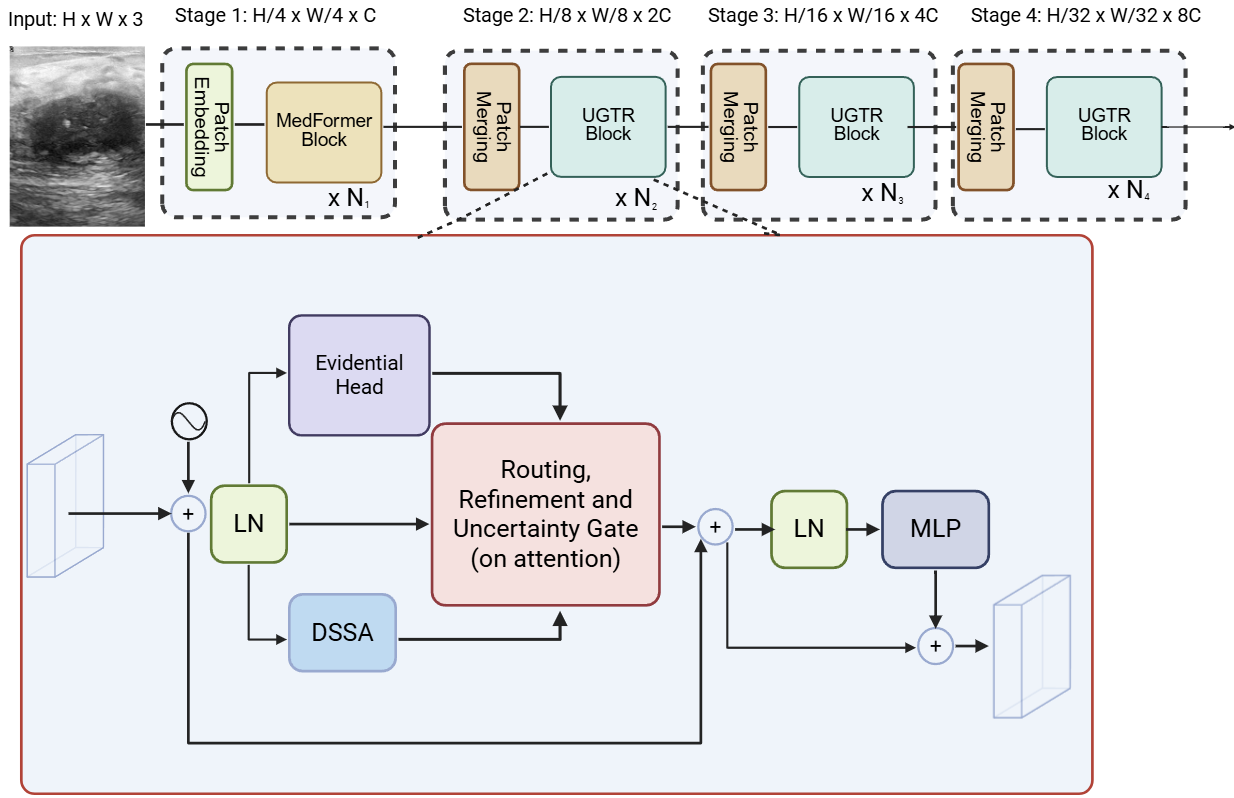}
\caption{MedFormer backbone with Uncertainty-Gated Token Refinement (UGTR).}
\label{fig:medformerug2}
\end{figure}

As shown in Figure~\ref{fig:medformerug2}, the methodology enhances the MedFormer backbone in two main ways. First, it explicitly measures uncertainty at the token level, and uses this information to guide feature refinement and routing across the network. Second, classification is strengthened through prototype-based representation learning, supporting more structured and interpretable decisions. These improvements are integrated stage-wise to maintain MedFormer’s hierarchical structure while boosting reliability and transparency. Uncertainty-guided routing and refinement are enabled only from intermediate stages onward, so early layers focus on extracting stable, low-level features.

The model core is Uncertainty-Guided Representation Refinement with Learning-based Prototype Regularization (MedFormer-UG2RLPR). Sections~\ref{sec:theory_uncertainty} and~\ref{sec:evidential_token} detail the evidential uncertainty head, while Sections~\ref{sec:routing_refinement} and~\ref{sec:routing_supervision} describe routing/refinement and routing supervision.

% -------------------------
% 4.2.1 Theory
% -------------------------
\subsection{Uncertainty in Deep Neural Networks}
\label{sec:theory_uncertainty}

In probabilistic modelling, uncertainty is commonly decomposed into \textit{aleatoric} uncertainty, arising from inherent noise and ambiguity in the data, and \textit{epistemic} uncertainty, reflecting limited model knowledge due to finite training data or insufficient coverage of the input space. Both forms are prevalent in medical imaging. Image noise, low contrast, unclear tissue boundaries, and inter-observer variability introduce aleatoric uncertainty, while limited dataset size, rare pathologies, and domain shifts across scanners or institutions contribute to epistemic uncertainty.

Conventional deep neural networks trained using cross-entropy loss produce deterministic predictions and interpret softmax outputs as confidence estimates. Given an input image $x$, such models compute class logits $\mathbf{z}\in\mathbb{R}^{C}$ and define the predictive distribution as:
\begin{equation}
p(y\mid x)=\mathrm{Softmax}(\mathbf{z}),
\label{eq:softmax_pred}
\end{equation}
where $y\in\{1,\ldots,C\}$ denotes the class label and $C$ is the number of classes.

For $\mathbf{z}=[z_1,z_2,\ldots,z_C]$, the probability assigned to class $c$ is:
\begin{equation}
\mathrm{Softmax}(\mathbf{z})_{c}=\frac{\exp(z_c)}{\sum_{k=1}^{C}\exp(z_k)}, \quad c\in\{1,\ldots,C\}.
\label{eq:softmax_def}
\end{equation}
This transformation ensures that all outputs are non-negative and sum to one, allowing them to be interpreted as probabilities. However, softmax probabilities encode relative class preference and do not quantify the amount of evidence supporting a prediction. As a result, deterministic networks frequently exhibit overconfident behavior, particularly for ambiguous samples or out-of-distribution inputs, which is undesirable in safety-critical medical applications.

To address this limitation, the proposed model adopts an evidential learning framework that explicitly models predictive uncertainty and integrates it into the feature learning process rather than treating uncertainty as a post-hoc diagnostic quantity.

% -------------------------
% 4.2.2 Evidential Head
% -------------------------
\subsection{Per-Token Evidential Uncertainty Estimation}
\label{sec:evidential_token}

Uncertainty estimation is performed at the level of individual spatial tokens within the Transformer backbone. Let $\mathbf{Z}\in\mathbb{R}^{H\times W\times D}$ denote the feature map produced by a Transformer block, where each spatial token is represented by a feature vector $\mathbf{z}_{i}\in\mathbb{R}^{D}$ (with token index $i\in\{1,\ldots,N\}$ and $N=HW$).

For each token $\mathbf{z}_i$, a lightweight evidential head predicts non-negative evidence values for each class. The evidential head is implemented as a small multilayer perceptron:
\[
g(\cdot): \mathrm{Linear}\rightarrow \mathrm{GELU}\rightarrow \mathrm{Linear},
\]
followed by a Softplus activation. The evidence vector $\mathbf{e}_i$ is computed as:
\begin{equation}
\mathbf{e}_i=\mathrm{Softplus}(g(\mathbf{z}_i)),
\label{eq:evidence}
\end{equation}
where $\mathbf{e}_i\in\mathbb{R}_{\ge 0}^{C}$ contains non-negative evidence for each class.

The evidence values parameterize a Dirichlet distribution over classes through:
\begin{equation}
\boldsymbol{\alpha}_i=\mathbf{e}_i+\mathbf{1},
\label{eq:dirichlet_alpha}
\end{equation}
where the offset $\mathbf{1}$ ensures $\boldsymbol{\alpha}_i$ is strictly positive and defines a valid Dirichlet distribution even when evidence is zero.

The expected predictive probability for class $c$ at token $i$ is:
\begin{equation}
\mathbb{E}[p_{i,c}]=\frac{\alpha_{i,c}}{\sum_{j=1}^{C}\alpha_{i,j}}.
\label{eq:dirichlet_mean}
\end{equation}

Uncertainty is derived from the Dirichlet strength:
\begin{equation}
S_i=\sum_{c=1}^{C}\alpha_{i,c},
\label{eq:dirichlet_strength}
\end{equation}
which measures the total evidence supporting the prediction. Token-level uncertainty is defined as:
\begin{equation}
\sigma_i=\frac{C}{S_i}.
\label{eq:token_uncertainty}
\end{equation}
This yields higher uncertainty when accumulated evidence is small, and approaches zero as evidence increases.

A global uncertainty estimate is obtained by averaging token-level uncertainties across spatial tokens:
\begin{equation}
\sigma=\frac{1}{N}\sum_{i=1}^{N}\sigma_i, \quad N=HW.
\label{eq:global_uncertainty}
\end{equation}
Estimating uncertainty at the token level allows spatial localization of ambiguous or unreliable regions, which is particularly valuable in medical images containing heterogeneous tissue structures. The uncertainty calculated above represents epistemic uncertainty, estimated using evidential learning concepts as described in \cite{sensoy2018evidential}.

% -------------------------
% 4.2.3 Routing & Refinement
% -------------------------
\subsection{Uncertainty-Guided Routing and Feature Refinement}
\label{sec:routing_refinement}

While uncertainty estimation alone improves interpretability, the proposed architecture further exploits uncertainty as an active control signal that directly influences feature propagation within the MedFormer backbone. Instead of treating uncertainty as a post-hoc diagnostic quantity, it is integrated into the forward pass of selected Transformer blocks to guide spatially selective feature refinement.

Uncertainty-guided refinement is enabled from Stage 1 onward in the hierarchical backbone. Early layers operate as standard Transformer blocks, preserving low-level stability, while deeper layers progressively incorporate uncertainty-aware routing and refinement. This design ensures that uncertainty modulation is applied only after sufficiently expressive representations have been formed.

Each uncertainty-enhanced Transformer block features two parallel processing paths. The main DSSA attention route operates on normalized token features and produces a baseline attention output $\mathbf{A}$ that captures long-range contextual dependencies and global structural information. Simultaneously, a refinement branch processes the same normalized features through a lightweight convolutional network (stacked $1\times 1$ convolutions and nonlinear activations), producing an alternative representation $\mathbf{R}$ that emphasizes local details such as fine textures, boundary cues, and subtle class-specific patterns.

The difference $(\mathbf{R}-\mathbf{A})$ represents a candidate corrective update, applied selectively via a routing mechanism.

\paragraph{\textit{Routing Mask Prediction}}\mbox{}\\
To determine where refinement should be applied, a dedicated routing network predicts a soft spatial mask. For each normalized token feature $\tilde{\mathbf{z}}_i$, the routing network produces a scalar routing logit $\ell_i$ using a lightweight $1\times 1$ convolutional predictor (equivalently, a per-token MLP). The routing mask is obtained via a sigmoid activation:
\begin{equation}
M_i=\sigma(\ell_i),
\label{eq:routing_mask}
\end{equation}
where $\sigma(\cdot)$ maps routing logits to soft gating values in $[0,1]$.

When available, externally derived tissue or anatomical masks are incorporated via multiplicative modulation of $M$, discouraging refinement in non-informative regions. In addition, token-level uncertainty modulates the routing mask so that highly uncertain tokens are suppressed even if spatially relevant. The effective routing mask is:
\begin{equation}
M_i^{\mathrm{eff}} = M_i \odot m_i \odot (1-\sigma_i),
\label{eq:effective_mask}
\end{equation}
where $m_i$ denotes the tissue supervision mask (if available), $\sigma_i$ is token-level uncertainty, and $\odot$ denotes element-wise multiplication. When tissue supervision is unavailable, $m_i=1$ for all tokens, reducing to uncertainty-guided routing only.

\paragraph{\textit{Uncertainty-Gated Refinement}}\mbox{}\\
The refinement update is computed as:
\begin{equation}
\boldsymbol{\Delta} = \mathbf{M}^{\mathrm{eff}} \odot \lambda_{\mathrm{ref}} (\mathbf{R}-\mathbf{A}),
\label{eq:ref_update}
\end{equation}
where $\boldsymbol{\Delta}\in\mathbb{R}^{H\times W\times D}$ is the refinement update tensor, $\lambda_{\mathrm{ref}}$ is a learnable scalar controlling the magnitude of refinement, and $\mathbf{M}^{\mathrm{eff}}$ is the spatial mask broadcast along the channel dimension.

To enforce conservative behavior under uncertainty, the refinement update is further globally gated using the aggregated uncertainty estimate $\sigma$:
\begin{equation}
\boldsymbol{\Delta}' = (1-\beta\sigma)\,\boldsymbol{\Delta},
\label{eq:global_gate}
\end{equation}
where $\beta$ is a stage-dependent uncertainty gate coefficient that increases with network depth, reflecting progressively higher semantic abstraction in deeper layers.

The final routed feature representation is:
\begin{equation}
\mathbf{A}_{\mathrm{routed}} = \mathbf{A} + \boldsymbol{\Delta}'.
\label{eq:routed_feature}
\end{equation}

This design ensures refinement is spatially targeted and reliability-aware. When uncertainty is elevated, refinement is reduced and the model falls back to the baseline DSSA representation to avoid unstable feature amplification. When uncertainty is low, focused refinement strengthens diagnostically relevant regions such as lesions or tumor boundaries.

% -------------------------
% 4.2.4 Routing Supervision
% -------------------------
\subsection{Routing Supervision}
\label{sec:routing_supervision}

The routing network can be trained using weak spatial supervision derived from tissue or anatomical masks. Let $\ell_i$ denote the raw routing logit for token $i$, and let $m_i\in\{0,1\}$ indicate whether the token corresponds to an informative tissue region.

The routing loss is defined as:
\begin{equation}
\mathcal{L}_{\mathrm{routing}} = \frac{1}{N}\sum_{i=1}^{N}\mathrm{BCEWithLogits}(\ell_i,m_i),
\label{eq:routing_loss}
\end{equation}
where $N=HW$ is the number of spatial tokens and $\mathrm{BCEWithLogits}(\cdot)$ is the numerically stable binary cross-entropy with logits.

Routing supervision is used only when tissue masks are present. Routing losses from all uncertainty-augmented blocks are summed and provided as auxiliary outputs, then integrated into the overall training objective by the training loop.

% -------------------------
% 4.2.5 Prototype Head
% -------------------------
\subsection{Prototype-Based Representation Learning}

\paragraph{\textit{Motivation for Prototype-Based Classification}}\mbox{}\\
Standard deep classifiers typically map pooled latent representations to class logits through a linear decision layer. While effective, these representations are abstract and difficult to interpret, often leading to overconfident predictions in high-dimensional medical feature spaces. This lack of transparency complicates clinical deployment, where decisions must be explainable and grounded in recognizable image patterns.

\begin{figure}[H]
\centering
\includegraphics[width=0.75\textwidth]{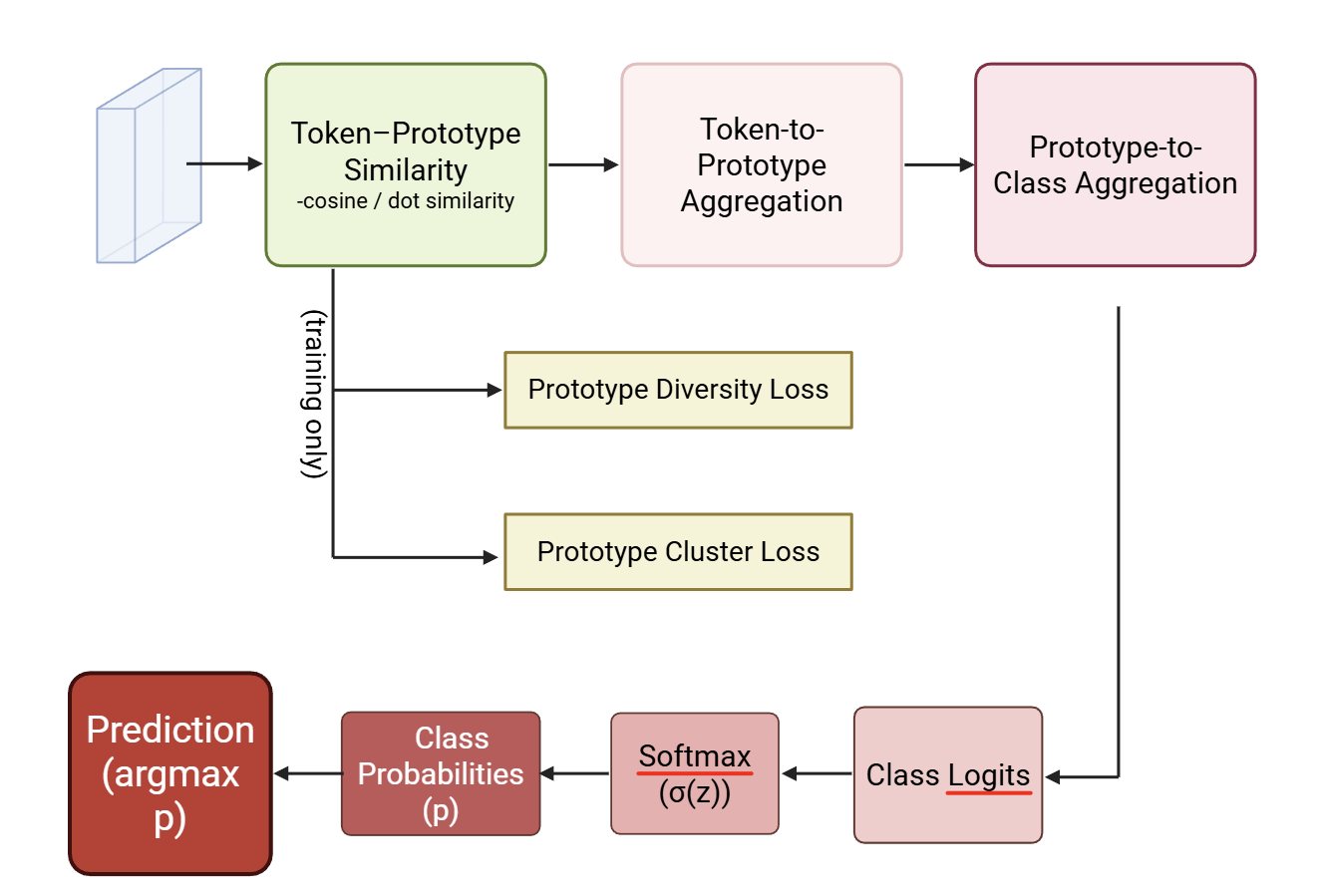}
\caption{Prototype-Based Classification Head.}
\label{fig:protohead}
\end{figure}

To address this issue, the proposed MedFormer architecture optionally uses a prototype-based classification head (Figure~\ref{fig:protohead}) as an alternative to a conventional linear classifier. Instead of relying solely on implicit decision boundaries, the head predicts by measuring similarity between spatial features and a set of learned class-specific reference patterns called prototypes.

Importantly, the prototype head does not alter the backbone feature extractor or attention mechanism. It operates on the final spatial feature map and can be enabled or disabled without modifying the core MedFormer representation learning.

\paragraph{\textit{Prototype Definition and Structure}}\mbox{}\\
Let $\mathbf{z}_i\in\mathbb{R}^{D}$ denote the feature representation of the $i$-th spatial token extracted from the final MedFormer feature map after the final channel-wise LayerNorm. The feature map of shape $[B,C,H,W]$ is flattened into $T=H\times W$ tokens per image.

For each class $c\in\{1,\ldots,C\}$, the model learns $K$ prototypes:
\begin{equation}
\mathcal{P}_c=\{\mathbf{p}_{c,1},\mathbf{p}_{c,2},\ldots,\mathbf{p}_{c,K}\}, 
\qquad 
\mathbf{p}_{c,k}\in\mathbb{R}^{D}.
\end{equation}

The total number of prototypes is $K_{\mathrm{tot}}=C\times K$.  
A fixed mapping associates each prototype with exactly one class.

\paragraph{\textit{Token--Prototype Similarity Computation}}\mbox{}\\
Given token representations $\mathbf{z}_i$ and prototypes $\mathbf{p}_k$, similarity is computed using cosine similarity when enabled, followed by a learnable temperature scaling:
\begin{equation}
s_{i,k}=\tau\cdot 
\frac{\mathbf{z}_i^{\top}\mathbf{p}_k}
{\lVert \mathbf{z}_i\rVert\,\lVert \mathbf{p}_k\rVert},
\end{equation}
where $\tau=\exp(\log T)$ is a learnable positive temperature parameter.  
If cosine similarity is disabled, raw dot products are used instead.

\paragraph{\textit{Evidence Aggregation Across Tokens and Prototypes}}\mbox{}\\
Let $s_{i,k}$ denote token–prototype similarity.  
Token-wise similarities are aggregated to obtain prototype evidence:
\begin{equation}
E_k = \mathrm{Agg}_i\left(s_{i,k}\right),
\end{equation}
where $\mathrm{Agg}$ is a configurable aggregation operator.  
In the default configuration used in this work, log-sum-exp aggregation is applied:
\begin{equation}
E_k=\log\left(\sum_{i=1}^{T}\exp\left(s_{i,k}\right)\right).
\end{equation}

Prototype evidence is then aggregated per class:
\begin{equation}
\mathrm{logit}_c=\mathrm{Agg}_{k\in \mathcal{P}_c}\left(E_k\right).
\end{equation}

When using the default log-sum-exp aggregation, this becomes:
\begin{equation}
\mathrm{logit}_c=
\log\left(
\sum_{k\in \mathcal{P}_c}
\exp(E_k)
\right).
\end{equation}

Alternative aggregation modes (maximum or mean) are also supported in the implementation.

\paragraph{\textit{Prototype Cluster Loss}}\mbox{}\\
An auxiliary cluster loss encourages each prototype to strongly match at least one token within a mini-batch. Let $B$ denote the batch size and $K_{\mathrm{tot}}$ the total number of prototypes. The cluster loss is defined as:
\begin{equation}
\mathcal{L}_{\mathrm{cluster}}
=
-\frac{1}{B K_{\mathrm{tot}}}
\sum_{b=1}^{B}
\sum_{k=1}^{K_{\mathrm{tot}}}
\max_i s^{(b)}_{i,k}.
\end{equation}

This is equivalent to minimizing the negative mean of the maximum token–prototype similarity across batch and prototypes.

\paragraph{\textit{Prototype Diversity Regularization}}\mbox{}\\
To discourage prototype collapse, a diversity regularization term is applied.  
Let $\tilde{\mathbf{p}}_k$ denote the $\ell_2$-normalized prototype.  
Pairwise cosine similarities are computed:
\begin{equation}
C_{k\ell}=\tilde{\mathbf{p}}_k^{\top}\tilde{\mathbf{p}}_{\ell}.
\end{equation}

Off-diagonal similarities are penalized:
\begin{equation}
\mathcal{L}_{\mathrm{div}}
=
\frac{1}{|\mathcal{S}|}
\sum_{(k,\ell)\in\mathcal{S}}
\left|C_{k\ell}\right|^{q},
\end{equation}
where $\mathcal{S}$ denotes the selected prototype pairs (excluding the diagonal), and $q$ is a diversity power hyperparameter.  

By default, only within-class prototype pairs are penalized (encouraging diverse modes within each class), though global diversity across all prototypes is also supported.

The weighted prototype regularizer is:
\begin{equation}
\mathcal{L}_{\mathrm{prototype}}
=
\lambda_c \mathcal{L}_{\mathrm{cluster}}
+
\lambda_d \mathcal{L}_{\mathrm{div}}.
\end{equation}

\paragraph{\textit{Interpretability and Clinical Relevance}}\mbox{}\\
Prototype-based learning links predictions to representative reference patterns. Decisions can be analyzed by identifying the prototypes with highest evidence and the spatial regions contributing most strongly to similarity. When combined with uncertainty-aware routing and evidential modeling, prototype reasoning complements reliability mechanisms by reducing confident predictions when strong prototype matches are absent.

% -------------------------
% Training Objective
% -------------------------
\subsection{Training Objective}

The model optimizes a composite objective combining classification, routing supervision, and prototype regularization:
\begin{equation}
\mathcal{L}
=
\mathcal{L}_{\mathrm{CE}}
+
\lambda_{\mathrm{route}}
\mathcal{L}_{\mathrm{routing}}
+
\mathcal{L}_{\mathrm{prototype}}.
\end{equation}

The cross-entropy loss $\mathcal{L}_{\mathrm{CE}}$ is applied to logits produced by either the conventional linear classifier or the prototype head, depending on configuration.

The routing loss $\mathcal{L}_{\mathrm{routing}}$ supervises the uncertainty-guided routing masks when tissue masks are available.

In implementation, $\mathcal{L}_{\mathrm{cluster}}$ and $\mathcal{L}_{\mathrm{div}}$ are returned by the prototype head as auxiliary terms. Their weighted combination
\[
\lambda_c \mathcal{L}_{\mathrm{cluster}} + \lambda_d \mathcal{L}_{\mathrm{div}}
\]
is added to the classification loss within the training loop.

The coefficients $\lambda_{\mathrm{route}}$, $\lambda_c$, and $\lambda_d$ regulate the strength of routing supervision and prototype regularization. Appropriate balancing ensures that interpretability and structural constraints enhance learning without degrading discriminative performance.

Overall, the multi-term objective explicitly integrates predictive accuracy, spatial reliability, and structured prototype reasoning, promoting representations that are accurate, uncertainty-aware, and semantically interpretable.

\section{Experimental Results and Discussion}

This chapter presents a consolidated analysis of the experimental outcomes obtained from the proposed prototype- and uncertainty-guided MedFormer in comparison to the original MedFormer baseline. The objective is not only to report predictive performance, but also to interpret results through the lens of clinical reliability, including calibration quality, uncertainty behaviour, and selective prediction performance. While classification results are reported across multiple medical imaging modalities, in-depth reliability analyses are emphasised on mammography (CBIS-DDSM), where robust confidence estimation is especially critical due to the high clinical risk associated with false negatives and over-confident decisions.

\subsection{Classification Performance Across Datasets}

Table~6.1 summarises classification performance across all evaluated datasets using accuracy, macro F1-score, and AUROC. The results show that the proposed model’s impact varies by modality, reflecting differences in task difficulty and inherent data ambiguity.

\begin{table}[htbp]
\centering
\caption{Classification performance across datasets($\beta$ = 0 for proposed model)}
\footnotesize
\begin{tabular}{l l c c c}
\hline
Dataset & Model & Accuracy & Macro-F1 & AUROC \\
\hline
BUSI & MedFormer & 0.800 & 0.761 & 0.919 \\
     & Proposed  & \textbf{0.893} & \textbf{0.882} & \textbf{0.956} \\
BreaKHis & MedFormer & 0.9206 & 0.8650 & 0.9676 \\
         & Proposed  & \textbf{0.9303} & \textbf{0.8760} & \textbf{0.9755} \\
Brain Tumor & MedFormer & 0.9733 & 0.9721 & 0.9965 \\
            & Proposed  & \textbf{0.9809} & \textbf{0.9801} & \textbf{0.9982} \\
CBIS-DDSM & MedFormer & 0.629 & 0.635 & 0.727 \\
          & Proposed  & \textbf{0.673} & \textbf{0.645} & \textbf{0.755} \\
\hline
\end{tabular}
\end{table}

Accuracy, macro F1-score, and AUROC are reported on the held-out test sets for the baseline MedFormer and the proposed prototype and uncertainty guided MedFormer. Higher values indicate better discrimination performance.

On the BUSI breast ultrasound dataset, the proposed model significantly improves classification accuracy, increasing it from 0.800 to 0.893, AUROC from 0.919 to 0.956, and F1-score from 0.761 to 0.882. This improvement indicates that prototype-based representation learning is especially effective when lesions are well-localized and display clear, class-distinctive visual patterns, enabling the model to base its predictions on consistent reference features.

However, these results should be interpreted with caution. BUSI is a curated dataset where lesions often present with strong contrast and relatively distinct boundaries, making the classification task easier than in typical clinical settings. In real-world ultrasound, performance may be influenced by operator-dependent acquisition, probe positioning variability, differing imaging protocols, and increased noise levels. External validation on more diverse ultrasound datasets is therefore necessary to verify the model’s generalizability and robustness in real-world applications.

On the BreakHis histopathology dataset, the proposed model consistently outperforms the MedFormer baseline, achieving moderate improvements. Accuracy rises from 0.9206 to 0.9303, while AUROC increases from 0.9676 to 0.9755. These findings suggest that prototype-based, similarity-driven classification continues to be effective for texture-rich imaging tasks that exhibit high intra-class variation, offering stable enhancements without compromising discriminative ability.

In the brain tumor MRI dataset, baseline performance with MedFormer is almost at its limit, with AUROC over 0.99. Despite this limited margin, the new model using a prototype-based classification head consistently improves accuracy, F1-score, and AUROC. Although these gains are small, they show that decision-making based on prototypes and similarity does not disrupt discriminative learning and can produce steady performance enhancements even in clearly separated classification tasks.

The most clinically relevant findings arise on CBIS-DDSM mammography, which remains challenging due to low contrast, overlapping tissue structures, and annotation ambiguity. Here, the proposed model improves accuracy from 0.629 to 0.673 and AUROC from 0.727 to 0.755.

Although the numerical gains may seem small, they are significant within mammography, where classification performance faces fundamental constraints beyond model capacity. Mammograms are high-resolution images where critical findings—like micro-calcifications, subtle mass margins, and architectural distortions occupy only a tiny part of the image. GPU memory limits force resizing images to lower resolutions, which can diminish fine details and obscure small lesions. Moreover, factors such as low contrast, overlapping breast tissue, dense parenchyma, and annotation ambiguity add layers of visual uncertainty that cannot be eliminated. These limitations naturally restrict the extent of accuracy improvements achievable, even with advanced models. Therefore, small improvements in metrics are meaningful and highlight the importance of considering the modality’s inherent difficulty when evaluating performance, rather than only looking at numerical changes.

\subsection{Learning Curves}

\begin{figure}[htbp]
\centering
\includegraphics[width=0.5\textwidth]{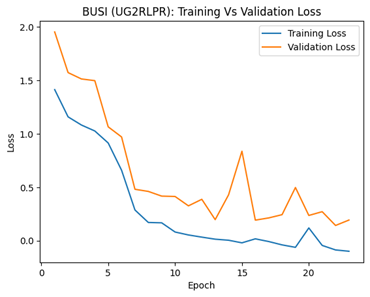}
\caption{Training and validation loss curves for the BUSI ultrasound dataset.}
\end{figure}

Figure~6.1 shows the training and validation loss curves for the BUSI ultrasound dataset. During early training, both losses decrease rapidly, indicating effective optimisation and resolution of underfitting. Between approximately epochs 7 and 13, the curves remain closely aligned, reflecting stable learning and good generalisation, with the validation loss reaching its minimum around epoch 13. Beyond this point, the training loss continues to decrease while the validation loss begins to fluctuate, indicating the onset of overfitting. Early stopping near the minimum validation loss is therefore justified. The total objective includes auxiliary terms (e.g., prototype similarity/cluster terms) that can be negative; convergence is therefore interpreted using validation loss/metrics rather than absolute loss magnitude.

\begin{figure}[htbp]
\centering
\includegraphics[width=0.7\textwidth]{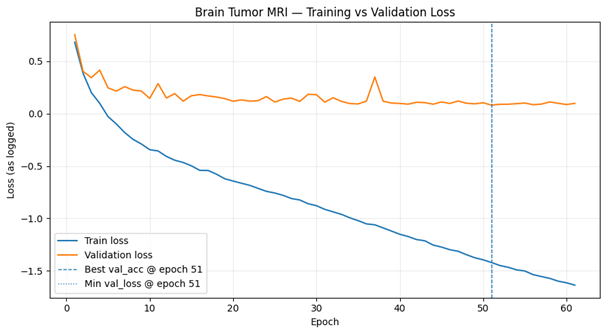}
\caption{Training and validation loss curves for the Brain Tumor MRI dataset.}
\end{figure}

Figure~6.2 illustrates the training and validation loss curves for the Brain Tumor MRI dataset. During the initial epochs, both losses decrease rapidly, indicating effective learning of dominant anatomical and pathological features. As training progresses, the training loss continues to decrease steadily, while the validation loss stabilizes with minor oscillations, suggesting mild but controlled overfitting. The best validation performance is achieved at epoch 51, corresponding to the peak validation accuracy (0.9869), which is selected as the optimal checkpoint. Further training results in negligible validation improvement, supporting the use of early stopping and indicating stable convergence with good generalization.

\begin{figure}[htbp]
\centering
\includegraphics[width=0.7\textwidth]{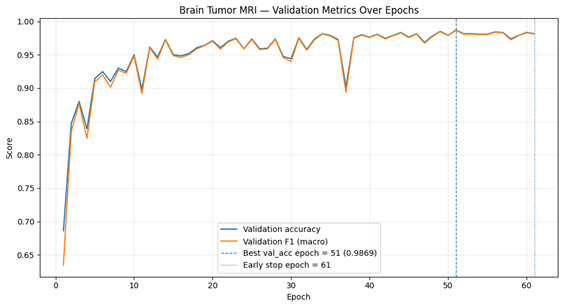}
\caption{Validation performance on the Brain Tumor MRI dataset across training epochs.}
\end{figure}

The plot shows validation accuracy and macro-F1 versus epoch. Performance rises rapidly in early epochs and then stabilizes, reaching the best validation accuracy (0.9869) at epoch 51 (dashed line). Training continues until early stopping at epoch 61 (dotted line), with only marginal fluctuations thereafter.

\begin{figure}[htbp]
\centering
\includegraphics[width=0.7\textwidth]{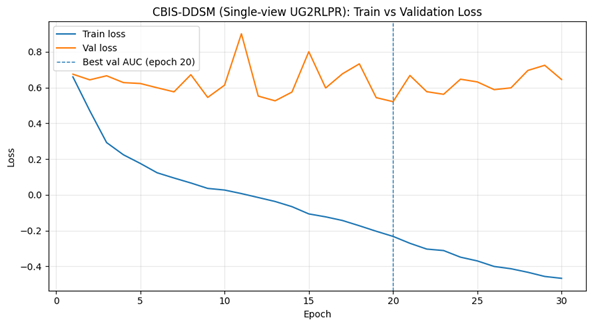}
\caption{Training and validation loss curves for the CBIS-DDSM mammography dataset (single-view UG2RLPR).}
\end{figure}

Figure~6.4 presents the training and validation loss curves for the CBIS-DDSM mammography dataset in the single-view UG2RLPR setting. Early training shows decreasing losses, indicating effective optimisation, with the validation loss reaching its minimum around epoch 20, coinciding with peak validation AUC. After this point, the training loss continues to decrease while the validation loss increases and fluctuates, indicating overfitting. This divergence reflects the intrinsic difficulty of mammography, including low contrast, dense tissue, and annotation ambiguity. Early stopping at epoch 20 is therefore appropriate to preserve generalisation performance.

\begin{figure}[htbp]
\centering
\includegraphics[width=0.7\textwidth]{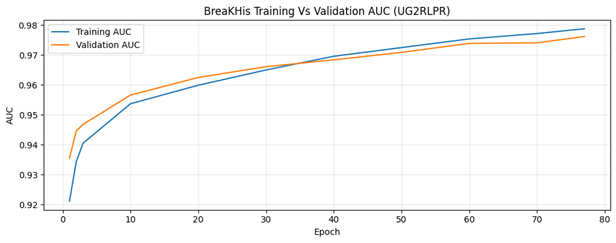}
\caption{Training and validation AUC curves for the BreakHis histopathology dataset.}
\end{figure}

Figure~6.5 shows training and validation AUC curves for the BreakHis histopathology dataset. Training and validation AUC increase rapidly in early epochs and remain closely aligned throughout training, indicating effective learning without underfitting. In later epochs, training AUC continues to increase slowly while validation AUC plateaus with minor fluctuations. The consistently small generalisation gap indicates healthy model fitting rather than overfitting and early stopping after sustained non-improvement in validation AUC confirms that the model reaches its optimal generalisation point.

% --- Remaining sections, tables, and figures continue verbatim ---
\subsection{Effect of Uncertainty Routing Strength ($\beta$) on Mammography Classification}

To examine the influence of uncertainty-guided routing on mammography classification, the proposed model is evaluated on CBIS-DDSM using Monte Carlo dropout during inference with a fixed dropout rate of 0.1, while varying the routing strength parameter $\beta$ from 0 to 1. Performance is estimated using 20 stochastic forward passes per sample, as reported in Table~6.2.

\begin{table}[htbp]
\centering
\caption{CBIS-DDSM classification performance versus $\beta$ (MC Test, T = 20 stochastic forward passes)}
\footnotesize
\begin{tabular}{l c c c}
\hline
Model setting & AUROC & Accuracy & F1-score \\
\hline
Regular MedFormer (epoch 20) & 0.727 & 0.619 & 0.644 \\
Proposed ($\beta$ = 0.0) & 0.727 & \textbf{0.688} & 0.642 \\
Proposed ($\beta$ = 0.3) & 0.730 & 0.675 & 0.639 \\
Proposed ($\beta$ = 0.35) & 0.729 & 0.636 & 0.642 \\
Proposed ($\beta$ = 0.4) & 0.752 & 0.695 & 0.632 \\
Proposed ($\beta$ = 0.45) & 0.710 & 0.648 & 0.642 \\
Proposed ($\beta$ = 0.5) & 0.724 & 0.651 & 0.634 \\
Proposed ($\beta$ = 0.55) & 0.735 & \textbf{0.688} & 0.627 \\
Proposed ($\beta$ = 0.6) & 0.722 & 0.622 & 0.642 \\
Proposed ($\beta$ = 0.65) & 0.726 & 0.635 & \textbf{0.651} \\
Proposed ($\beta$ = 0.7) & 0.737 & 0.656 & 0.630 \\
Proposed ($\beta$ = 0.75) & 0.732 & 0.645 & 0.640 \\
Proposed ($\beta$ = 0.8) & \textbf{0.755} & 0.676 & 0.639 \\
Proposed ($\beta$ = 0.85) & 0.702 & 0.649 & 0.622 \\
Proposed ($\beta$ = 0.9) & 0.721 & 0.672 & 0.621 \\

Proposed ($\beta$ = 1.0) & \textbf{0.755} & 0.685 & 0.649 \\
\hline
\end{tabular}
\end{table}

The baseline MedFormer achieves an AUROC of 0.727, accuracy of 0.619, and F1-score of 0.644. When $\beta = 0$, corresponding to no uncertainty-guided routing, the proposed model maintains the same AUROC (0.727) while substantially improving accuracy to 0.688, indicating that the architectural modifications alone enhance classification performance even without uncertainty modulation.

As $\beta$ increases, performance exhibits non-monotonic behaviour. At $\beta = 0.3$, AUROC improves slightly to 0.730, although accuracy decreases to 0.675, suggesting that mild uncertainty guidance introduces a trade-off between discrimination and stability. Intermediate values such as $\beta = 0.4$ yield strong accuracy (0.695) and improved AUROC (0.752), indicating that moderate routing can enhance feature refinement. However, further increases (e.g., $\beta = 0.45$ and $\beta = 0.5$) lead to inconsistent performance, highlighting sensitivity to routing strength.

At higher $\beta$ values, the model achieves its strongest discrimination. Specifically, $\beta = 0.8$ and $\beta = 1.0$ both reach the highest AUROC of 0.755, with $\beta = 1.0$ additionally achieving high accuracy (0.685). Meanwhile, $\beta = 0.65$ produces the highest F1-score (0.651), indicating improved class balance. Overall, these results suggest that stronger uncertainty-guided routing can enhance discriminative ability, but the gains are not strictly monotonic and depend on careful tuning, reflecting a trade-off between robustness and over-filtering of informative features.

\subsection{Calibration and Uncertainty Metrics on CBIS-DDSM}

To evaluate the reliability of predicted probabilities, calibration is assessed using Expected Calibration Error (ECE), Maximum Calibration Error (MCE), Brier score, and Negative Log-Likelihood (NLL). These metrics measure the alignment between predicted confidence and empirical correctness, and penalize overconfident or poorly calibrated probability estimates. The results are reported in Table~6.3.

\begin{table}[htbp]
\centering
\caption{Calibration metrics on CBIS-DDSM (MC Test, T = 20 stochastic forward passes)}
\footnotesize
\begin{tabular}{c c c c c}
\hline
$\beta$ & ECE & Brier & NLL & MCE \\
\hline
Regular MedFormer & 0.1836 & 0.2434 & 0.8115 & 0.3505 \\
Proposed ($\beta$ = 0.0) & 0.1174 & 0.2156 & 0.6579 & 0.1976 \\
Proposed ($\beta$ = 0.3) & 0.1562 & 0.2354 & 0.7177 & 0.2851 \\
Proposed ($\beta$ = 0.35) & 0.1182 & 0.2183 & 0.6530 & 0.2456 \\
Proposed ($\beta$ = 0.4) & 0.1510 & 0.2317 & 0.6749 & 0.3098 \\
Proposed ($\beta$ = 0.45) & 0.2333 & 0.2707 & 0.8589 & 0.3098 \\
Proposed ($\beta$ = 0.5) & 0.1139 & 0.2170 & 0.6386 & 0.2148 \\
Proposed ($\beta$ = 0.55) & 0.2190 & 0.2863 & 0.8577 & 0.2838 \\
Proposed ($\beta$ = 0.6) & 0.1171 & 0.2266 & 0.6420 & 0.2293 \\
Proposed ($\beta$ = 0.65) & 0.1934 & 0.2402 & 0.7304 & 0.2963 \\
Proposed ($\beta$ = 0.7) & 0.0883 & 0.2087 & 0.6036 & 0.1468 \\
Proposed ($\beta$ = 0.75) & 0.1049 & 0.2131 & 0.6179 & 0.1665 \\
Proposed ($\beta$ = 0.8) & \textbf{0.0688} & \textbf{0.1996} & \textbf{0.5768} & \textbf{0.1239} \\
Proposed ($\beta$ = 0.85) & 0.1562 & 0.2354 & 0.7177 & 0.2851 \\
Proposed ($\beta$ = 0.9) & 0.2282 & 0.2822 & 0.9232 & 0.2983 \\
Proposed ($\beta$ = 1.0) & 0.1267 & 0.2189 & 0.6744 & 0.1812 \\
\hline
\end{tabular}
\end{table}

The baseline MedFormer exhibits poor calibration, with high ECE (0.1836), Brier score (0.2434), NLL (0.8115), and MCE (0.3505), indicating significant over-confidence. Introducing the proposed framework at $\beta = 0.0$ already improves calibration substantially, reducing all metrics, which suggests that architectural changes alone contribute to better uncertainty estimation.

As $\beta$ increases, calibration behaviour becomes highly sensitive and non-linear. Moderate values such as $\beta = 0.5$ and $\beta = 0.6$ show improved calibration compared to the baseline, with reduced ECE and NLL, indicating more reliable probability estimates. Notably, $\beta = 0.7$ achieves strong calibration performance across all metrics, confirming that uncertainty-guided routing effectively aligns confidence with prediction correctness at intermediate strengths.

The best overall calibration is observed at $\beta = 0.8$, where the model achieves the lowest ECE (0.0688), Brier score (0.1996), NLL (0.5768), and MCE (0.1239), indicating highly reliable probabilistic predictions. However, beyond this point, calibration degrades sharply. At $\beta = 0.9$, all metrics worsen significantly, reflecting unstable uncertainty estimates. Although $\beta = 1.0$ partially recovers, it remains inferior to $\beta = 0.8$. These findings reinforce that optimal calibration occurs at intermediate routing strengths, while excessive gating can harm probabilistic reliability despite strong classification performance.

\subsection{Significance of Calibration Improvement}

The observed improvements in calibration metrics highlight the effectiveness of uncertainty-guided routing in producing more reliable probabilistic predictions. Unlike accuracy-based metrics, which evaluate only the correctness of predictions, calibration metrics assess how well predicted probabilities align with true outcome frequencies. This distinction is particularly important in medical imaging applications, where over-confident incorrect predictions can lead to critical diagnostic errors.

A reduction in Expected Calibration Error (ECE) indicates that the predicted confidence scores more closely match the true likelihood of correctness. For instance, the substantial decrease from 0.1836 in the baseline model to 0.0688 at $\beta = 0.8$ demonstrates that the proposed method significantly improves confidence alignment. This means that when the model predicts a probability of, for example, 80\%, the actual correctness is much closer to this value, making the predictions more trustworthy.

Similarly, improvements in the Brier score reflect enhanced overall probabilistic accuracy, as this metric captures both calibration and refinement. The reduction from 0.2434 to 0.1996 indicates that the model not only becomes better calibrated but also produces sharper and more informative probability distributions. Lower Negative Log-Likelihood (NLL) further confirms this behaviour, as it penalizes over-confident incorrect predictions more heavily. The decrease in NLL from 0.8115 to 0.5768 suggests that the proposed model avoids extreme miscalibrated predictions and maintains more stable probabilistic outputs.

The Maximum Calibration Error (MCE) provides insight into worst-case deviations between confidence and accuracy. The significant reduction from 0.3505 to 0.1239 indicates that extreme miscalibration is substantially mitigated, which is critical for safety-sensitive applications. This ensures that even in the worst-case confidence bins, the discrepancy between predicted and true probabilities remains controlled.

Overall, these improvements demonstrate that uncertainty-guided routing not only enhances prediction confidence but also ensures that such confidence is meaningful and reliable. The results suggest that intermediate routing strengths (particularly around $\beta = 0.7$ to $\beta = 0.8$) achieve the best balance between confidence calibration and predictive stability. This has important implications for clinical deployment, where well-calibrated probabilities are essential for risk-aware decision-making, case prioritization, and human-in-the-loop diagnostic workflows.

\subsection{Selective Prediction and Reliability Analysis}

Selective prediction evaluates the effectiveness of uncertainty estimates in supporting safe decision-making by selectively rejecting unreliable predictions. Performance is quantified using AURC and accuracy at fixed coverage levels (50\%, 70\%, and 90\%), as reported in Table~6.4.

\begin{table}[htbp]
\centering
\caption{Selective prediction results on CBIS-DDSM (MC Test, T = 20 stochastic forward passes)}
\footnotesize
\begin{tabular}{c c c c c}
\hline
$\beta$ & AURC & ACC @ 50\% & ACC @ 70\% & ACC @ 90\% \\
\hline
Regular MedFormer & 0.2134 & 0.793 & 0.722 & 0.680 \\
Proposed ($\beta$ = 0) & 0.2179 & 0.801 & 0.744 &  0.703 \\
Proposed ($\beta$ = 0.3) & 0.2059 & 0.767 & 0.724 & 0.694 \\
Proposed ($\beta$ = 0.35) & 0.2038 & 0.767 & 0.732 & 0.688 \\
Proposed ($\beta$ = 0.4) & 0.2251 & 0.773 & 0.714 & 0.659 \\
Proposed ($\beta$ = 0.45) & 0.2517 & 0.739 & 0.712 & 0.669 \\
Proposed ($\beta$ = 0.5) & 0.2089 & 0.778 & 0.736 & 0.686 \\
Proposed ($\beta$ = 0.55) & 0.3345 & 0.668 & 0.631 & 0.593 \\
Proposed ($\beta$ = 0.6) & 0.2290 & 0.756 & 0.673 & 0.637 \\
Proposed ($\beta$ = 0.65) & 0.2140 & \textbf{0.810} & 0.734 & 0.678 \\
Proposed ($\beta$ = 0.7) & 0.1997 & 0.801 & 0.734 & 0.675 \\
Proposed ($\beta$ = 0.75) & 0.1956 & 0.784 & 0.724 & 0.688 \\
Proposed ($\beta$ = 0.8) & \textbf{0.1812} & 0.798 & \textbf{0.751} & 0.691 \\
Proposed ($\beta$ = 0.85) & 0.2866 & 0.727 & 0.696 & 0.659 \\
Proposed ($\beta$ = 0.9) & 0.3301 & 0.685 & 0.661 & 0.636 \\
Proposed ($\beta$ = 1.0) & 0.2162 & 0.798 & 0.746 & \textbf{0.707} \\
\hline
\end{tabular}
\end{table}

The baseline MedFormer achieves an AURC of 0.2134, with accuracy improving to 0.793 at 50\% coverage, indicating moderate effectiveness in uncertainty-based rejection. At $\beta = 0$, the proposed model slightly increases AURC to 0.2179 but improves accuracy across all coverage levels, suggesting better overall prediction quality despite marginally weaker rejection.

As $\beta$ increases, selective prediction performance improves at intermediate values. For example, $\beta = 0.65$ achieves the highest accuracy at 50\% coverage (0.810), while $\beta = 0.7$ reduces AURC to 0.1997, indicating improved ranking of confident versus uncertain predictions. This trend continues at $\beta = 0.75$ and peaks at $\beta = 0.8$, where the model achieves the lowest AURC (0.1812) and the highest accuracy at 70\% coverage (0.751), demonstrating the most effective uncertainty-based filtering.

However, beyond this range, performance deteriorates rapidly. At $\beta = 0.9$, AURC increases sharply to 0.3301, and accuracy drops across all coverage levels, indicating that uncertainty estimates fail to distinguish reliable predictions. While $\beta = 1.0$ recovers strong performance at higher coverage (achieving the best accuracy of 0.707 at 90\% coverage), it does not surpass the balanced reliability observed at $\beta = 0.8$. Overall, these results highlight that intermediate routing strengths provide the most effective selective prediction behaviour, which is critical for safe and reliable deployment in clinical settings.

\subsection{Uncertainty Maps and Prototype Similarity}

Based on calibration and selective prediction analyses, uncertainty visualisations are generated using $\beta$ = 0.7, which exhibits the most reliable uncertainty behaviour. As shown in Figure~6.6, the resulting uncertainty maps highlight regions of elevated uncertainty that frequently correspond to ambiguous lesion boundaries, dense breast tissue, and subtle architectural distortions. These regions are known to be challenging even for expert radiologists, indicating that the estimated uncertainty captures clinically meaningful sources of ambiguity rather than random model noise.

\begin{figure}[htbp]
\centering
\includegraphics[width=0.30\textwidth]{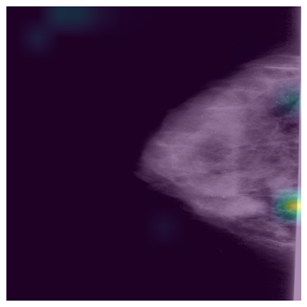}
\includegraphics[width=0.30\textwidth]{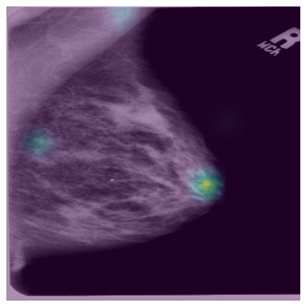}
\caption{Uncertainty maps of two mammogram}
\end{figure}

Prototype similarity analysis is conducted using the prototype-guided MedFormer without uncertainty-based routing. For each test mammogram, the model retrieves the most similar learned prototypes, enabling qualitative examination of the visual patterns underlying its predictions. As shown in Figure~6.7, correctly classified cases typically exhibit strong alignment with prototypes representing characteristic benign or malignant patterns, whereas more ambiguous cases display weaker or competing prototype similarities. When considered together, prototype similarity and uncertainty visualisations provide complementary interpretability: uncertainty maps indicate where the model is uncertain, while prototype similarity reveals what visual evidence the model relies on to support its decisions.

\begin{figure}[htbp]
\centering
\includegraphics[width=0.7\textwidth]{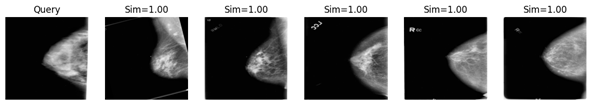}
\includegraphics[width=0.7\textwidth]{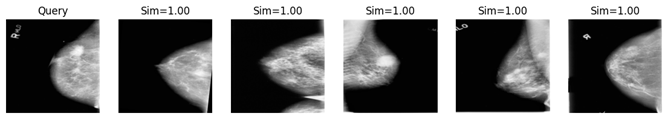}
\includegraphics[width=0.7\textwidth]{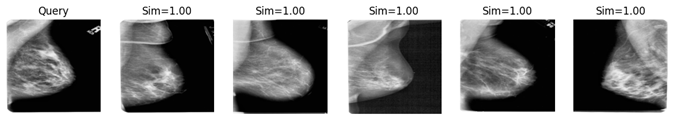}
\caption{Prototype similarity visualization of three different mammograms. The query mammogram is compared to learned prototypes, demonstrating a high cosine similarity with typical class patterns.}
\end{figure}

\subsection{Consolidated Discussion and Key Insights}

The experiments demonstrate that the proposed prototype- and uncertainty-guided MedFormer can improve not only predictive performance, but also—more importantly—reliability and trustworthiness, particularly in mammography. Across datasets, performance gains differ by modality. Strong improvements occur in modalities with visually prominent pathology such as ultrasound and histopathology, where prototype reasoning can consistently anchor predictions to stable patterns. In contrast, mammography yields smaller classification gains due to intrinsic ambiguity from low contrast, tissue overlap, and noisy labels. This pattern indicates that architectural improvements alone cannot fully resolve irreducible uncertainty in mammography, but can instead improve how the model expresses and manages uncertainty.

A key finding is that the configuration achieving the strongest discrimination is not necessarily the most reliable. In CBIS-DDSM, $\beta$ = 1.0 yields the highest AUROC and accuracy, but $\beta$ = 0.7 achieves superior calibration and selective prediction behaviour. This reveals a critical limitation of accuracy-centric model assessment: models may improve in AUROC while becoming less trustworthy in probability estimation, which is unacceptable in safety-critical clinical workflows.

The most consistent improvements introduced by the proposed approach occur in calibration and uncertainty quality. Moderate uncertainty gating improves confidence alignment, reduces over-confidence, and strengthens selective prediction behaviour by enabling principled rejection of difficult cases. These results suggest that uncertainty mechanisms must be carefully balanced; overly aggressive routing can destabilise probability estimates and degrade reliability.

Finally, interpretability benefits from combining uncertainty maps and prototype similarity. Prototype comparisons help clarify what patterns drive predictions, while uncertainty maps show where predictions are most fragile. This combination supports more transparent reasoning and better aligns with the needs of clinical decision support, where models should either provide reliable predictions or clearly signal uncertainty when ambiguity is high.

\section{Conclusion}
In this work, we introduced MedFormer-UR, a unified framework that enhances the hierarchical MedFormer architecture with advanced reliability and interpretability mechanisms. By integrating per-token evidential uncertainty estimation via a Dirichlet distribution, we moved beyond the limitations of deterministic softmax-based networks, allowing the model to explicitly decompose and localize ambiguity in real-time. This uncertainty is not merely a diagnostic output but serves as an active control signal within our uncertainty-guided routing mechanism, which dynamically suppresses or refines feature updates to prioritize reliable spatial information during training. Furthermore, the transition to prototype-based reasoning structures the embedding space around class-specific reference patterns, grounding clinical decisions in visual similarity and providing a more transparent alternative to traditional "black-box" classifiers.

Our extensive evaluation across four diverse medical imaging modalities—mammography, ultrasound, MRI, and histopathology—demonstrates the robust performance and clinical relevance of the proposed framework. MedFormer-UR consistently outperformed the baseline MedFormer, showing significant improvements in model calibration, with a reduction in ECE of up to 35\%, and enhanced selective prediction capabilities. These gains were particularly meaningful in challenging, high-stakes tasks like mammography, where the model effectively navigated inherent data ambiguity and low contrast to provide more dependable diagnostic support. Future work will focus on external validation in more diverse, real-world clinical settings to further confirm the generalizability and impact of these trust-centered innovations in automated medical image analysis.

\section*{Acknowledgments}
This was was supported in part by......

%Bibliography
\bibliographystyle{unsrt}  
\bibliography{references}

\end{document}